\documentclass[a4paper]{JHEP3}
\usepackage{epsfig}
\usepackage{graphicx}
\usepackage{amssymb}
\usepackage{latexsym}

\newcommand{\be}{\begin{equation}}
\newcommand{\ee}{\end{equation}}
\newcommand{\bea}{\begin{eqnarray}}
\newcommand{\eea}{\end{eqnarray}}

\title{Optimal weak value measurements: Pure states.}
\author{N.D. Hari Dass\footnote{dass@tifrh.res.in}
\\  Visiting Professor, TIFR-TCIS, Hyderabad 500075.\\}
\author{Rajath Krishna R\footnote{rajathkrishnar@gmail.com}\\
 St. Xavier's College, Mumbai 400001. } 
\author{Sai Smruti Samantaray\footnote{smruti.samantaray89@gmail.com}\\
Department of Physics,IIT-M, Chennai 600036.}

\abstract{We apply the notion of \emph{optimality} of measurements for state determination(tomography) as originally given by
Wootters and Fields to \emph{weak value tomography} of \emph{pure states}. They defined measurements to be optimal if they 'minimised'
the effects of statistical errors.
For technical reasons they actually maximised the state averaged information, precisely quantified as the negative logarithm
of 'error volume'. In this paper we optimise both the state averaged information as well as error volumes. 
We prove, for Hilbert spaces of arbitrary
(finite) dimensionality, that varieties of weak value measurements are optimal when the post-selected bases are \emph{mutually unbiased} 
with respect to the eigenvectors of the observable being measured. We prove a number of important results about the geometry of state spaces 
when
expressed through the weak values as coordinates. We derive an expression for the Ka\"ehler potential for the N-dimensional case with the help
of which we give an exact treatment of the arbitrary-spin case.
}


\begin{document}

\section{Introduction}
State determination through measurements, also called \emph{Tomography}, is very important in Quantum Mechanics. It is also important in 
Classical mechanics, but it is considerably more nuanced and involved in quantum theory. We recall here a very deep characterisation of
states in general given by Dirac\cite{diracstate}, which can be invoked even in the absence of any Hilbert space structure; according to him, \emph{states}
are the embodiment of the collection of all possible measurement outcomes. Fortunately, both in classical 
as well as in quantum mechanics( described by finite-dimensional Hilbert spaces), it is
sufficient to collect measurement outcomes for finite number of observales constituting a \emph{complete set}. For a N-dimensional quantum
system, the state is generically represented by a Hermitian, unit trace density matrix requiring $N^2-1$ real numbers for its complete specification. The outcomes of any observable, at least in the projective measurement scheme, are N eigenvalues along with N probabilities for them. 
Only the probabilities carry information about the state and since their sum must equal unity, each measurement yields $N-1$ real parameters. 
Thus, in order to obtain the required $N^2-1$ real parameters, one has to measure $N+1$ linearly
independent observables. In the case of qubits, for example, one needs three such measurements. These could be, say, measurements of $S_x,S_y,S_z$, or, equally well of ${\vec S}\cdot{\vec n}_i$ along three non-collinear directions ${\vec n}_i$.

While both sets are equally good in terms of state determination, they are not so, according to
Wootters and Fields \cite{woottersmub}, in terms of their accuracies in state determination. Errors are inevitable in measurements. One could
for example take the variance as a measure of this error(reduced by the usual statistical factor of $M^{-1/2}$ with M being the number of
measurements). 
Interpreting the expectation values of the complete set of operators $O^\alpha$ as \emph{coordinates} of the space of states, the measurement 
errors can be taken to be the extents of an \emph{error parallelepiped} centred at the point representing the 
state in which the measurements were carried out. Then, according to Wootters and Fields, a tomography is optimal when basically the geometrical volume 
of the parellelepiped is the smallest. 
Variance $\Delta O$ in general depends on both O as well as the state $\rho$ of the system. The volume element
additionally depends on the metric of the state space. Thus in general the error volume has a state dependence. 
It is, however, quite meaningless to optimise the error volume for a given state. This is because in general the
criterion for optimality(in the Wootters-Fields case, 
the ${\vec n}_i$); in tomography, the state is a priori \emph{unknown}. Because of this, it
is impossible to choose the suitable observables before hand. The best that can be done is to use the 
\emph{expected errors} for random choices of the state. In other words, one should only work with state averaged error volumes. Wootters and Fields found it convenient to work with
\emph{state averaged 
tomographic information} and they found that it is the greatest 
when the ${\vec n}_i$ are mutually orthogonal as in $S_x,S_y,S_z$. The important property they highlight for this choice can be expressed
as, for $\alpha\,\ne\,\beta$,
\be
\label{eq:qubit-mub}
|\langle\uparrow^\beta|\uparrow^\alpha\rangle|^2\,=
|\langle\uparrow^\beta|\downarrow^\alpha\rangle|^2\,=
|\langle\downarrow^\beta|\downarrow^\alpha\rangle|^2\,=
\,\frac{1}{2}
\ee
Here $|\uparrow^\alpha\rangle,|\downarrow^\alpha\rangle$ refer to the eigenvectors of the observable $O^\alpha$. The three sets of
eigenvectors in this case are said to form \emph{Mutually Unbiased Bases}(MUB) in the sense that eigenvectors of one operator have 
equal probabilties of outcome if that operator is measured on any of the other eigenvectors. They were first introduced by Schwinger
\cite{schwingermub} 
who called such bases \emph{complimentary}. For the N-dimensional cases, eqn.(\ref{eq:qubit-mub}) generalises to
\be
\label{eq:Ndim-mub}
|\langle\,k^\alpha|j^\beta\rangle|^2\,=\,\frac{1}{N}\quad\quad \alpha\,\ne\,\beta
\quad\quad\,|\langle\,k^\alpha|j^\alpha\rangle|^2\,=\,\delta_{jk}
\ee
Thus, the central result of Wootters and Fields \cite{woottersmub} is that measurements with the complete set $O^\alpha$ will be optimal
if their eigenstates $|k^\alpha\rangle$ form a MUB. Mutually unbiased bases have subsequently been seen to play a fundamental role in
diverse contexts \cite{mubdiverse}. Adamson and Steinberg \cite{mubproof} have experimentally vindicated the Wootters-Fields result.
There has been a paradigm shift in quantum measurements with the so called \emph{Weak Measurements} proposed by Aharonov et al.\cite{aharorig}.
There are
many ways to qualitatively understand how weak measurements work; one such is to replace the narrow pointer states
used for the initial state of an apparatus in a von Neumann model of projective(also called strong) measurements with a broad and coherent superposition of such 
narrow pointer states\cite{threeweakndh}. We prefer this description as it counters the commonly held view that weak measurements require a weaker \emph{interaction} 
between the system and the apparatus as compared to the \emph{strong} measurements. It is the ratio of
the displacement of the mean pointer position of the apparatus to the width of the apparatus state that
is relevant. We also make a clear distinction between such 
weak measurements and the so called 
\emph{Weak Value Measurements} which are weak measurements followed by \emph{Post-selection} realised through strong measurements. It has
been pointed out that weak values are special cases of the Dirac-Kirkwood quasiprobabilities \cite{dirackirkwood}.

Quite remarkably, weak value measurements offer a radically different and novel means of tomography. While such \emph{weak tomography} have
been proposed for both pure as well as mixed states \cite{lutomo1,lutomo2,swutomo}, we shall restrict ourselves for the moment to 
only pure states. Then, unlike the
standard tomography based on complete sets of observables, weak tomography yields complete information for pure states by making measurements
on the $N\,-1\,$ \emph{Projection operators} for \emph{one} given single observable. The $N\,-1\,$ independent \emph{complex} weak values are directly 
measurable. This is in the sense that the post-measurement state of the apparatus can generically be described by a gaussian centred around
the weak value. But the width of such gaussians are also very very large.
Unlike a general density matrix which requires $N^2\,-\,1$ real parameters for its complete description, a pure state requires only $2N\,-\,2$
real, or $N\,-\,1$ complex parameters. Therefore, weak value measurements are naturally suited for pure state tomography in arbitrary
dimensions. The weak values can be taken as the complex coordinates for the state space. It turns out to be a sterographic projection of
the state space. This has been experimentally verified by Kobayashi, Nonaka and Shikano 
for optical vortex-beams \cite{weakstereo}. For a detailed account of weak measurements and numerous references see \cite{nori,swutomo}

In this work we have analysed the Wootters-Fields optimality criteria for weak value
tomography of pure states. We point out without going to details that other precision criteria have also been studied \cite{zhouhall,vallone,pangfisher}.
We state our main results right away and provide all relevant
details in the following. In standard tomography one can compare measurements done with different complete sets for their optimality.
Since in the case of weak tomography, the observable for measurement is fixed(we consider all
different projection operators as just different aspects of this single observable), one will have to compare different choices of 
post-selection for optimality. Our principal result is that weak value measurements are optimal when post-selected states are mutually 
unbiased wrt to the eigenfunctions of the observable under measurement. We show this by explicit calculations for spin-1/2, spin-1 and
spin-3/2 cases. Then we prove the result for \emph{arbitrary} spins.  Computing error volumes, and subsequently minimising them 
requires the \emph{metric} on the space of states. The state space for pure states is actually a \emph{Projective Hilbert Space} and
these are known to be not only \emph{Complex Manifolds} but also the so called \emph{Ka\"ehler Manifolds}. Metrics of such spaces are completely
fixed in terms of a single scalar function called the \emph{Ka\"ehler Potential}\cite{kaehlerbook}. For the above three cases the metric components are calculated
explicitly and the respective Ka\"ehler potentials determined. The Ka\"ehler potential for the arbitrary spin case is then deduced by induction.

It should be mentioned that such special post-selection states were already used in \cite{lutomo1,lutomo2,swutomo}. In \cite{lutomo1} a post-selected states were chosen that were not only mutually unbiased wrt to the eigenstates
of the measured observable, even their relative phases were constant. 
This
ensured that the weak values were actually proportional to the pure state to be determined. In \cite{swutomo} too MUB's were used more as
means of simplifying analysis rather than as fundamental necessities. In fact tomography can be accomplished without any of these simplifications i.e when the post-selected states are not MUB's wrt eigenstates of observables. In this paper we establish the fundamental result that it is 
only when post-selected states are MUB's the measurements are the most precise in the sense of the error volumes being the smallest.
\section{\normalsize Optimal Weak Tomography of a Spin-1/2 System}
In this section we give a review of the work done by Hari Dass \cite{threeweakndh} on optimal weak measurement of a spin-1/2 system. 
If $|{\pm}\rangle$ are eigenvectors of, say, $S_z$,
 any pure state can be written as
\be
\label{eq:qubitpure}
|{\psi}\rangle\,=\alpha_+ |{+}\rangle\,+\alpha_- |{-}\rangle\,\quad |\alpha_+|^2\,+\,|\alpha_-|^2\,=\,1
\ee
The weak values for the measurements of the projectors $\Pi_{\pm}\,=\,|{{\pm}}\rangle\langle{{\pm}}|$,
with post-selected state $|{b}\rangle$ are \cite{aharorig}, with 
$b_{\pm}=\langle{b}|{{\pm}}\rangle$, and $\phi_0$ the \emph{phase} of $\langle{b}|\psi\rangle$,
\be
\label{eq:weakvalues-qubit}
w_{\pm}\,=\,\frac{\langle{b}|{{\pm}}\rangle\langle{{\pm}}|{\psi}\rangle}{\langle{b}|{\psi}\rangle}\rightarrow
\alpha_{\pm}= \frac{w_{\pm}/b_{\pm}}{\sqrt{|\frac{w_+}{b_+}|^2+|\frac{w_-}{b_-}|^2}}\,e^{i\phi_0}\quad\quad w_+\,+\,w_-\,=\,1 
\ee
Thus exactly two independent real parameters are left for the
parametrisation of the qubit state. The density matrix for qubits(both pure and mixed) can be represented as
\be\label{eq:rhoqubit}
\rho=\frac{I}{2}+\langle S_x \rangle \sigma_x+\langle S_y \rangle\sigma_y+\langle S_z \rangle\sigma_z
\ee
It should be noted that density matrices can be represented in terms of complete sets of observables irrespective of whether the
actual tomography is based on that complete set or not. In terms of $\alpha_{\pm}$, the expectation values occurring above are given by
\be
\label{eq:qubitexpectations}
\langle\,S_x\,\rangle\,=\, Re\,\alpha_+^*\,\alpha_-\quad
\langle\,S_y\,\rangle\,=\, Im\,\alpha_+^*\,\alpha_-\quad
\langle\,S_z\,\rangle\,=\, \frac{1}{2}\,\,(|\alpha_+|^2\,-\,|\alpha_-|^2)\,
\ee
While $\alpha_{\pm}$ are merely parametrisations of the density matrices, tomography lies in their determination through
measurements. In weak tomography, $w_{\pm}$ are directly measured and are related to the parameters in the density
matrix, by eqn.(\ref{eq:weakvalues-qubit}).
We define the distance function(hence the metric) on the space of states(density matrices) by 
\be
\label{eq:metriconrho}
dl^2\,=2\,Tr\,d\rho d\rho
\ee
This coincides with the Fubini-Study metric \cite{crell} for pure states, but differs from it for mixed states. Nevertheless it is an acceptable metric even
for mixed states with the natural isometries inherited from quantum theory.
We get the line element to be 
\be
\label{eq:qubitmetric}
dl^2\,=\,4\,[(d\langle S_x \rangle)^2\,+\,(d\langle S_y \rangle)^2\,+\,(d\langle S_z \rangle)^2]
\ee
Let us now consider an intermediate step of rewrting $\alpha_{\pm}$ as
\be
\label{eq:qubitzpm}
\alpha_{\pm}\,=\,\frac{z_{\pm}}{\sqrt{|z_+|^2\,+\,|z_-|^2}}
\ee
Indeed eqn.(\ref{eq:weakvalues-qubit}) without the constraint $w_+\,+\,w_-\,=\,1$ is of this form with 
$z_{\pm}\,=\,\frac{w_{\pm}}{b_{\pm}}\,e^{i\phi_0}$. Without any constraints, $z_{\pm}$
are {\bf four} real parameters, two too many for the qubit-state space. But scaling both z's by a common complex number $\lambda$ 
changes the $\alpha$'s by a common phase and hence the state, consequently the line element, remain
unchanged. 
So the redundancy in z's is two which gets exactly 
removed by the one complex constraint. So one can first work out the metric in terms of $z_{\pm}$ and then impose the constraint.

Since the line-element does not change, $g_{z_+z_+}=g_{z_-z_-}=0$. The
constraint, rewritten as
\be
\label{eq:zconstraint}
w_+\,+\,w_-\,=\,1\rightarrow b_+z_+\,+\,b_-z_-=1\rightarrow b_+dz_+\,+\,b_-dz_-=0
\ee
while reducing the number of parameters by 2 does not mix $dz_{\pm}$ with $d{\bar z}_{\pm}$. 
Hence $g_{z_+z_+},g_{z_-z_-}$ continue to be zero even after
the constraint.
Explicit coordinates for such projective spaces are usually chosen by fixing one of the z's to be
a constant, say, unity. In the weak value coordinates case this is done by the weak value constraint 
$w_+\,+\,w_-\,=\,1$.
Though algebraically more elaborate, this is a natural choice dictated by the constraint on weak values. But this choice 
introduces  explicit $b_i$ dependences into the otherwise purely geometrical entities such as the metric and Ka\"ehler potential etc. 
and the precise $b_i$ dependences are critical as they determine the optimality criteria.
The line element for qubit pure state space in weak-value coordinates can be workred out(after some tedious algebra) to be(in what follows we shall use $G_{ij}$ for the \emph{complex} metric and $G\,=\,det G_{ij}$ for its determinant , and likewise, $g_{ij}$ for
the metric in real coordinates and $g\,=\,det g_{ij}$ for the determinant of that metric.
\be
\label{eq:qubitweakmetric}
dl^2\,=\,\frac{4}{|b_+|^2|b_-|^2\big ( |\frac{w_+}{b_+}|^2+|\frac{w_-}{b_-}|^2\big )^2}
 dw_+d{\bar{w_+}}\,=\,G_{w_+{\bar w}_+}\,dw_+\,d{\bar w}_+
\ee
This metric is \emph{conformal} (see also \cite{korotkov}, for other conformal features of qubit state-space). From our general arguments this is just a reflection of the projective nature of this state space. In the qubit case, with just one complex coordinate, this is all there is to it. For the higher-dimensional cases, all the complex metric components of
the type $G_{w_iw_j}$ have to vanish again. 
The line element in terms of the real coordinates $x=Re w_+$ and $y=Im w_+$ is given by
\be
\label{eq:qubitrealmetric}
dl^2\,=\,\frac{4|b_+|^2|b_-|^2(dx^2+dy^2)}{((x\,-\,|b_+|^2)^2+y^2+|b_+|^2|b_-|^2)^2}\,=\,g_{ij}\,dx^i\,dx^j
\ee  
It is well-known that these state-spaces are Ka\"ehler manifolds \cite{kaehlerbook} for which the nonvanishing components of the metric is given by
\be
\label{eq:kahlerpot}
G_{w_i{\bar w}_j}\,=\,\partial_{w_i}\,\partial_{{\bar w}_j}\,K(\{w_i,{\bar w}_j\})
\ee
$K$ is called the Ka\"ehler potential. It is straightforward to show that for the qubit case
($w_+\,+\,w_-\,=\,1$)
\be
\label{eq:qubitkahler}
K_{qubit}(w_+,{\bar w}_+) \,=\,4\,\log\,\{|\frac{w_+}{b_+}|^2\,+\,|\frac{w_-}{b_-}|^2\}
\ee
A very important relation to notice is that between the Ka\"ehler potential K and the determinant g(G) of the metric:
\be
\label{eq:KGrelnqubit}
g_2\,=\,det g_{ij}\,=\,\frac{16}{|b_+|^4|b_-|^4}\,e^{-K}
\ee
Such a relationship is basic to Ka\"ehler metrics \cite{kaehlerbook}. 
These
techniques and results are of wide generality.

The volume element $\sqrt{g_2}\,dxdy$, where $g\,=\,det\,g_{ij}$, is given by 
\be
\label{eq:qubitvol}
dV_2\,=\,\frac{4|b_+|^2|b_-|^2\,dxdy}{((x\,-\,|b_+|^2)^2+y^2+|b_+|^2|b_-|^2)^2}
\ee
A consistency check is to calculate the total area which should be \emph{independent} of the $b_i$. An important property of weak value measurements comes into play at this stage i.e the
weak values are \emph{unbounded}. Because of this the total area integral can be computed by shifting the x-variable and integrating over
both coordinates over $(-\infty,\infty)$. The answer one gets is $A\,=\,4\pi$, which is the area of a sphere, and is indeed independent of
the $b_i$! Thus the weak value coordinates provide
a sterographic projection of the sphere onto a plane \cite{weakstereo,korotkov}. 

Now we come to an evaluation of the error area. Again, subtle features of the weak value measurements become crucial. The measurement of 
$Re\,w_+$ is done, in the original scheme \cite{aharorig,swutomo,nori}, by momentum measurements. The post-measurement apparatus 
state in that case is a gaussian in momentum space centred around $2\,Re\,w$ but with a very very large width, say, $\Delta$. On the other hand, 
measurement of $Im\,w$ has to be done independently by position measurements. For the same initial apparatus state, the post-measurement
state is now a \emph{narrow} gaussian in position space of width $\simeq\,\frac{1}{\Delta}$, but centred around $2\,Im\,w\,\Delta^{-2}$.Thus the error in $Im\,w$
is also large i.e $\Delta$. Strictly speaking the variance in weak measurements is $\sqrt{\Delta^2\,+\,{\Delta_\psi\,S_z}^2}$ \cite{nori}, 
but the second term is totally negligible. For an ensemble of M measurements each, these are reduced by the usual $\sqrt{M}$ factor giving a
statistical error $\Delta_s$ that can be taken to be small enough to use the \emph{local} volume element( in what follows, we shall take the extents of the error volumes to be $2\Delta_s$ in each direction)
\be
\label{eq:qubiterrorarea}
\Delta V_2^{err}\,=\,\frac{16|b_+|^2|b_-|^2\,\Delta_s^2}{((x\,-\,|b_+|^2)^2+y^2+|b_+|^2|b_-|^2)^2}
\ee
It should be noted that when $\Delta_s$ is not small, one will have to express this as a rather complicated integral. But the most important
difference from the Wootters-Fields analysis is that the error $\Delta_s$
is state independent., whereas in their case the errors being variances in given state depend both on the choice of the observable as well
as on the state. On the other hand, the metrics on state space in their cases are essentially flat and do not depend on the state. But in the
weak tomographies the metric is state-dependent. 

One could have contemplated minimising $\Delta V_2^{err}$ itself wrt $|b_i|$ for a given state. Even for
a given state, changing $b_i$ would change $w_i$. In fact the relevant form of $\Delta V_2^{err}$ to
consider would be
\be
\label{eq:errorvolpsifixed}
\Delta V_2^{err}\,=\,\frac{16\,\Delta_s^2}{|b_+|^2\,|b_-|^2}\,|\langle b|\psi\rangle|^4
\ee
It is indeed possible to find the stationary points i.e $|b_{\pm}|\,=\,|\psi_{\mp}|$. But for tomography of an unknown state, there is no way to post-select accordingly. Therefore in both cases one has to consider state averaged error volumes before 
optimising them. The state average of any function $f(x,y)$ on the state space is given by
\be
\label{eq:stateaverages}
\langle\,f(x,y)\,\rangle\,=\,\frac{\int\,\sqrt{g}\,dx\,dy\,f(x,y)}{\int\,\sqrt{g}\,dx\,dy}
\ee
Carrying out the state averaging, one finds, for the qubit case,
\be
\label{eq:stateaverrqubit}
\langle \Delta V_2^{err} \rangle=\frac{16\Delta_s^2}{3\,|b_+|^2|b_-|^2}  
\ee
This state averaged error volume takes its minimal value when $|b_+|^2=|b_-|^2=\frac{1}{2}$(recall that $|b_+|^2\,+\,|b_-|^2\,=\,1$) 
i.e., the post-selected state should be mutually unbiased with respect to the eigenstates of the operator being measured. 

\section{\normalsize Optimal weak tomography of Spin-1 System}
If $|{i}\rangle$ are eigenvectors of $S_z$ with $i\,=\,+1,0,-1$,
any spin-1 pure state can be written as
\be
\label{eq:qutritstate}
|{\psi}\rangle=\sum_{i=1}^3\,\alpha_i\,|{i}\rangle
\ee

The weak values for the projectors $\Pi_i\,=\,|{i}\rangle\langle{i}|\,$ with post-selected state $|{b}\rangle$ are given by
\be
\label{eq:weakvalues-qutrit}
w_{i}\,=\,\frac{\langle{b}|{{i}}\rangle\langle{{i}}|{\psi}\rangle}{\langle{b}|{\psi}\rangle}\rightarrow
\alpha_{i}= \frac{w_{i}/b_{i}}{\sqrt{|\frac{w_+}{b_+}|^2+|\frac{w_-}{b_-}|^2}}\,e^{i\phi_0}\quad\quad w_+\,+\,w_0\,+\,w_-\,=\,1 
\ee
We express the 3x3 density matrix, requiring 8 real parameters for its full description, in terms of 
the complete set $T_i\,=\,\frac{\Lambda_i}{2},i=1,..,8$, where the $\Lambda_i$ are the
Gell-Mann matrices satisfying the algebra
\be
\label{eq:gellmannmatrices}
[\Lambda_i,\Lambda_j]\,=\,2\,i\,f_{ijk}\,\Lambda_k\quad\quad 
\{\Lambda_i,\Lambda_j\}\,=\,\frac{4}{3}\,\delta_{ij}\,+2\,d_{ijk}\,\Lambda_k\,\rightarrow\,Tr \Lambda_i\,\Lambda_j\,=\,2\,\delta_{ij}
\ee
The density matrix is then represented as
\be
\label{eq:rhoqutrit}
\rho\,=\,\frac{I}{3}\,+\,\langle{T_i}\rangle\,\Lambda_i\quad\quad \langle{T_i}\rangle\,=\,Tr\,\rho\,T_i
\ee
%
The metric on spin-1 state space(valid for both pure and mixed states) is, then,
\be
\label{eq:qutritmetric}
dl^2\,=\,4\,\sum_i\,d\langle{T_i}\rangle\cdot\,d\langle{T_i}\rangle
\ee
For the pure state of eqn.(\ref{eq:qutritstate}), one gets,
\bea
\label{eq:Lambdaexpectation}
\langle{T_1}\rangle\,&+&\,i\,\langle{T_2}\rangle=\,\alpha_+^*\alpha_0;
\langle{T_4}\rangle\,+\,i\,\langle{T_5}\rangle\,=\,\alpha_+^*\alpha_-;
\langle{T_6}\rangle\,+\,i\,\langle{T_7}\rangle\,=\,Im\,\alpha_+^*\alpha_-;\nonumber\\
\langle{T_3}\rangle\,&=&\,\frac{1}{2}\,(|\alpha_+|^2\,-\,|\alpha_0|^2);
\langle{T_8}\rangle\,=\,\frac{1}{2\sqrt{3}}\,(|\alpha_+|^2\,+\,|\alpha_0|^2\,-\,2\,\,|\alpha_-|^2)
\eea
Resulting in the metric 
\bea
\label{eq:qutritmetric2}
D&*&(dl)^2\,= 
\bigg [1-w_- -{\bar w}_{-} +\frac{w_{-} {\bar w}_{-}}{|b_-|^2}\bigg ]\frac{dw_+{d{\bar w}}_+}{|b_+|^2}+
\bigg [ \frac{{\bar w}_+{\bar b}_-}{{\bar b}_+}+\frac{w_-b_+}{b_-}-\frac{{\bar w}_+w_-}{{\bar b}_+b_-} \bigg ] 
\frac{dw_+{d{\bar w}}_-}{b_+{\bar b}_-}\nonumber\\
&+& 
\bigg [\frac{{\bar w}_-{\bar b}_+}{{\bar b}_-}+\frac{w_+b_-}{b_+}-\frac{{\bar w}_-w_+}{{\bar b}_-b_+} \bigg ]\frac{dw_-d{\bar w}_+}{b_-{\bar b}_+}+
  \bigg [1-w_+ -{\bar w}_{+} +\frac{w_{+} {\bar w}_{+}}{|b_+|^2}\bigg ]\frac{dw_-d{\bar w}_-}{|b_-|^2} 
\eea
Where
\be
\label{eq:qutritmetricdenom}
D\,=\,\frac{|b_0|^2}{4}\Big (\frac{|w_+|^2}{|b_+|^2}+\frac{|w_0|^2}{|b_0|^2}+\frac{|w_-|^2}{|b_-|^2}\Big )^2
\ee
We have checked these results by choosing another complete set constructed out of the 4 MUB's given in 
\cite{kurzynski}
(see also \cite{durtenglert}) by choosing two independent projectors from each of the 4 sets.
%
%
The metric is no longer conformal as in the qubit case but still satisfies the $g_{w_iw_j}\,=\,0$. 
It is
Ka\"ehler, 
\be\label{eq:qutritkahler}
G_{w_i{\bar w}_j}\,=\,\partial_{w_i}\,\partial_{{\bar w}_j}\,K_{qutrit}\quad\quad
K_{qutrit}\,=\,4\,\ln\,(\sum_i\,|\frac{w_i}{b_i}|^2)
\ee
Changing to real coordinates $w_+=x_1+ix_2,w_-=x_3+ix_4$ 
The determinant  $g_3$ of the real metric is given by
\be
\label{eq:detgqutrit}
g_3=\frac{256}{|b_+|^4|b_0|^4|b_-|^4 \Big (\frac{|w_+|^2}{|b_+|^2}+\frac{|w_0|^2}{|b_0|^2}+\frac{|w_-|^2}{|b_-|^2}\Big )^6}
\ee
Once again there is a direct relation between $K_{qutrit}$ and $g_3$:
\be
\label{eq:KGrelnqutrit}
g_3\,=\,det g_{ij}\,=\,\frac{16}{|b_+|^4|b_0|^4|b_-|^4}\,e^{-\frac{3K}{2}}
\ee
The volume element is given by $dV_3=\sqrt{g_3}dx_1dx_2dx_3dx_4$
The total volume of the state space turns out to be
$V_3\,= \int dV=8\pi^2$. 
Note that this is not the surface-volume of sphere in 5 dimensions! The pure state space is a sphere only for the qubit case.

The error volume is computed along similar lines as in the qubit case and turns out to be
\be
\label{eq:qutriterrorvol} 
\Delta V_3^{err}= \frac{256 \Delta_s^4 }{|b_+|^2|b_0|^2|b_-|^2 \Big (\frac{|w_+|^2}{|b_+|^2}+\frac{|w_0|^2}{|b_0|^2}+\frac{|w_-|^2}{|b_-|^2}\Big )^4} 
\ee
The state averaged error volume is
\be
\label{eq:averrorqutrit}
\langle \Delta V_3^{err}\rangle= \frac{128\Delta_s^4}{5(|b_+|^2|b_0|^2|b_-|^2)}\quad\quad |b_+|^2\,+\,|b_0|^2\,+\,|b_-|^2\,=\,1
\ee
The above expression is minimum when $|b_+|^2=|b_0|^2=|b_-|^2=\frac{1}{3}$ and thus the measurement is optimal when the post-selected states 
are mutually unbiased with respect to the eigenstates of the observable being measured. 
\section{\normalsize Generalisation to arbitrary spins.}

The forms of eqns.(\ref{eq:qubitkahler},\ref{eq:qutritkahler}) strongly suggest 
the Ka\"ehler Potential for the general case 
\be
\label{eq:kahlerpotgen}
K_{N}\,=\,4\,\ln\,(\sum_{i=1}^N\,|\frac{w_i}{b_i}|^2)\quad\quad \sum_{i=1}^N\,w_i\,=\,1\quad\quad \sum_{i=1}^N\,|b_i|^2\,=\,1
\ee
This can also be shown by
\emph{induction} 
as when $w_N$ is set to zero, one should recover the $N-1$-dimensional case and that the
Ka\"ehler potential is completely symmetric in the variables $z_i\,=\,\frac{w_i}{b_i}$. 
We have explicitly verified this for spin-3/2(qudit) case 
on using the SU(4) Gell-Mann matrices
\cite{qudit}.In fact, even for the general case, on taking the observables to be half the SU(N) 
Gell-Mann
matrices normalised according to $Tr\,\Lambda_i\,\Lambda_j\,=\,2\,\delta_{ij}$, the analogs of 
eqns.(\ref{eq:rhoqubit},\ref{eq:qubitmetric},\ref{eq:rhoqutrit},\ref{eq:qutritmetric}) all turn out to be of identical forms. 
Comparing with eqns.(\ref{eq:KGrelnqubit},\ref{eq:detgqutrit}), a suggestive generalisation to arbitrary spin case for the determinant $g_N$ of the metric in
real coordinates($w_i\,=\,x_i\,+\,i\,y_i$ for i=1,..,N-1) is
\be
\label{eq:detgN}
g_N\,=\,\frac{4^{2N-2}}{\prod_{i=1}^N\,|b_i|^4}\,\frac{1}{(\sum_{i=1}^N\,|\frac{w_i}{b_i}|^2)^{2N}}
\ee
Actually the Ka\"ehler potential has all the information one needs and it can be shown that eqn.(\ref{eq:detgN}) can be derived from
eqn.(\ref{eq:kahlerpotgen}). Once again we see a direct relation between $g_N,K_N$:
\be
\label{eq:KGrelngen}
g_N\,=\,det g_{ij}\,=\,\frac{4^{2N-2}}{\prod_{i=1}^N\,|b_i|^4}\,e^{-\frac{NK_N}{2}}
\ee
The volume element $dV_N$ in the general case is
\be
\label{eq:dVgen}
d\,V_N\,=\,\frac{4^{N-1}}{\prod_{i=1}^N\,|b_i|^2}\,\frac{1}{(\sum_{i=1}^N\,|\frac{w_i}{b_i}|^2)^{N}}\,\prod_{i=1}^{N-1}\,dx_i\,dy_i
\ee
The total volume of the state space is
\be
\label{eq:Vgen}
V_N\,=\,\int\,dV_N
\ee
The error volume in the general case is, likewise,
\be
\label{eq:dVerrgen}
\Delta\,V_N^{err}\,=\,\frac{4^{N-1}(2\Delta_s)^{2N-2}}{\prod_{i=1}^N\,|b_i|^2}\,\frac{1}{(\sum_{i=1}^N\,|\frac{w_i}{b_i}|^2)^{N}}
\ee
The state averaged error volume is calculated as before. In calculating both the total volume as well as state
averaged error volumes, one has to evaluate integrals of the type
\be
\label{eq:weakintegrals}
I_M\,=\,\int\,\prod_{i=1}^{N-1}\,dx_i\,dy_i\,\frac{1}{(\sum_{i=1}^N\,|\frac{w_i}{b_i}|^2)^M}
\ee
In the case of $V_N$, $M=N$, and in the case of $\langle V^{err}_N\rangle$, $M=2N$. Upon eliminating $w_N$ by $w_N\,=\,1-\sum_{i=1}^{N-1}\,w_i$, and
expressing in terms of real coordinates, the
denominator(without the power M) can be expressed as
\be
\label{eq:genden}
D\,=\,x^T\cdot {\cal M}\cdot x\,+\,y^T\,M\,y\,+\,c_N\,-\,2\,c_N\,{\tilde D}^T\cdot x
\ee
with ${\cal M}$ a symmetric N-1xN-1 matrix and ${\tilde D}$ a N-1 column vector given by
\be
\label{eq:MtildeD}
{\cal M}_{ij}\,=\,c_i\,\delta_{ij}\,+\,c_N\quad\quad {\tilde D}^T\,=\,\{1,1,\ldots,1\}\quad\quad c_i\,=\,|b_i|^{-2}
\ee
Satisfying
\be
\label{eq:weakintegral2}
det {\cal M}\,=\,\prod_{i=1}^N\,|b_i|^{-2}\quad\quad c_N\,-\,c_N^2\,{\tilde D}^T\cdot\,M^{-1}\,{\tilde D}\,=\,1
\ee
Thus
\be
\label{eq:weakintegrals2}
I_M\,=\,\prod_{i=1}^N\,|b_i|^2\,\Omega_p\,\int\,dR\,\frac {R^{2N\,-\,3}}{(R^2\,+\,1)^M}
\ee
where $\Omega_p\,=\,\frac{2\,\pi^{p/2}}{\Gamma(p/2)}$ is the solid angle in p dimensions.
On using the definite integrals
\be
\label{eq:defints}
\int_0^\infty\,dx\,\frac{x^{N-2}}{(1\,+\,x)^N}\,=\,\frac{1}{N\,-\,1}\quad\quad
\int_0^\infty\,dx\,\frac{x^{N-2}}{(1\,+\,x)^{2N}}\,=\,\frac{\Gamma(N\,-\,1)\,\Gamma(N\,+\,1)}{\Gamma(2\,N)}
\ee
the final results for $V_N$ and $\langle\Delta V_N^{err}\rangle$ are evaluated to be 
\be
\label{eq:resultsNdim}
V_N\,=\,\frac{4^{N\,-\,1}}{N\,-\,1}\,\frac{\pi^{N\,-\,1}}{\Gamma(N\,-\,1)}\quad\quad\,\langle \Delta V_N^{err} \rangle = \frac{4^{2\,N\,-2}\,\Delta_s^{2N-2}}{(|b_1|^2|b_2|^2...|b_N|^2)}\,\frac{\Gamma(N)\,\Gamma(N\,+\,1)}{\Gamma(2\,N)}
\ee

For optimal weak measurement, we have to minimize this error volume. $\langle\,\Delta V_N^{err}\,\rangle$ is smallest when 
$|b_1|^2=|b_2|^2=....=|b_N|^2=\frac{1}{N}$ and thus the measurement is optimal when the post-selected states are mutually unbiased wrt the eigenstates of the observable being measured. 
\section{Proof based on information}
Now we show how to prove this by maximising information as done in \cite{woottersmub}. Following them, 
the information is taken to be 
\be
\label{eq:weakinfo}
{\cal I}\,=\,-\ln\,\Delta V_N^{err}\,=\,-(2\,N\,-\,2)\,\ln\,2\,+\,\sum_{i=1}^N\,\ln\,|b_i|^2\,+\,N\,\ln\,\sum_{i=1}^N\,|\frac{w_i}{b_i}|^2
\ee
The state averaged information is, then,
\be
\label{eq:stateavweakinfo}
\langle\,{\cal I}\,\rangle\,=\,-\,(2N\,-\,2)\,\ln\,2\,-\,(2\,N\,-\,2)\,\ln\,2\Delta_s\,+\frac{N\,2^{2\,N\,-\,2}\Omega_{2\,N\,-\,2}}{V_N}\,{\tilde I}_N\,+\,\sum_{i=1}^N\,\ln\,|b_i|^2
\ee
Here ${\tilde I}$ stands for
\be
\label{eq:tildeI}
{\tilde I}_N\,=\,\int\,dR\,\frac{2\,R^{2\,N\,-\,3}\,\ln R}{(R^2\,+\,1)^N}
\ee
It is easy to see that maximising this is equivalent to minimising the state averaged error volume.
The main subtlety is that the log of the average need not equal the average of the log, but in our case their
difference is independent of $b_i$.
\section*{\normalsize Other weak tomography methods.}

As in any post-selection only a fraction $|\langle{b}|\psi\rangle|^2$ of the data is made use of, ways have been
suggested in \cite{lutomo1,swutomo,lutomo2} to overcome this. They consist in performing weak tomography with a
larger set of $|b_i\rangle$ even a complete set of such post-selected states. It is clear that our analyses can be
applied to each post-selected state and the general result that optimal measurements require $|b_i\rangle$ to be 
mutually unbiased wrt eigenstates of the observable continues to hold. 



In \cite{swutomo} it was shown that for weak tomography of a pure state it is sufficient to do measurements of a 
single projector $A_{\phi}=|{\phi}\rangle\langle{\phi}|$ but with a full bases of $|b_j\rangle$ 
where $|{\phi}\rangle$, is subject to
$\langle{b_j}|{\phi}\rangle\neq 0$, but otherwise arbitrary. For each $|b_j\rangle$ the measured (complex)weak values are
\be
\label{eq:swuphiweakvalue}
W_j\,=\,\frac{\langle{b_j}|\phi\rangle\langle\phi|\psi\rangle}{\langle{b_j}|\psi\rangle}\quad \sum_j\,|\langle\phi|{b_j}\rangle|^2\,=\,1
\ee
Unlike the weak values of earlier tomography 
\be
\label{eq:weaksumne1}
\sum_j\,W_j\,\ne\,1
\ee
However, we can introduce new complex values ${\tilde w}_j$
\be
\label{eq:newphivalues}
{\tilde w}_j\,=\,\frac{|\langle\phi|{b_j}\rangle|2}{W_j}\,=\,\frac{\langle\phi|b_j\rangle\langle{b_j}|\psi\rangle}{\langle\phi|\psi\rangle}
\ee
Formally, ${\tilde w}_j$ can be thought of as the N complex weak values one would obtain by measuring the 
projectors $|b_j\rangle\langle{b_j}|$ with $|\phi\rangle$ as the post-selected state.
The corollary of our results 
would be that $|\phi\rangle$ should be mutually unbiased to the basis $\{|b_j\rangle\}$. In other words, the
measurements are optimal when $|\langle\phi|b_j\rangle|^2\,=\,\frac{1}{N}$ for every j. Since for every system 
there always exist at least two sets of MUB \cite{durtenglert}, such optimal measurements can be realised in many ways.

\acknowledgments
RK thanks TCIS-Hyderabad for the hospitality during which this work was carried out. NDH thanks Justin Dressel 
for many enlightening discussions. 

\end{document}